\documentclass[twocolumn,showpacs,preprintnumbers,amsmath,amssymb]{revtex4}%
\usepackage{graphicx}
\usepackage{dcolumn}
\usepackage{bm}
\usepackage{times}
\usepackage{amsmath}
\usepackage{amsfonts}
\usepackage{amssymb}%
\providecommand{\U}[1]{\protect\rule{.1in}{.1in}}
\newcommand{\be}{\beta}

\def\be{\begin{equation}}
\def\ee{\end{equation}}
\begin{document}
\title{PERFECT FLUID DARK MATTER}
\author{F. Rahaman}
\email{farook_rahaman@yahoo.com}
\affiliation{Department of Mathematics, Jadavpur University, Kolkata 700 032, West Bengal, India}
\author{K. K. Nandi }
\email{kamalnandi1952@yahoo.co.in}
\affiliation{Department of Mathematics, North Bengal University, Siliguri - 734013, West
Bengal, India}
\author{A. Bhadra}
\email{aru_bhadra@yahoo.com}
\affiliation{High Energy and Cosmic Ray Research Centre, University of North Bengal,
Siliguri, WB 734013, India }
\author{M. Kalam}
\email{mehedikalm@yahoo.co.in}
\affiliation{Department of Physics, Netaji Nagar College for Women, Regent Estate,
Kolkata-700092, India}
\author{K. Chakraborty}
\email{kchakraborty28@yahoo.com}
\affiliation{Department of Physics, Government Training College, Hooghly - 712103, West
Bengal, India}
\date{\today}

\begin{abstract}
\noindent Taking the flat rotation curve as input and treating the
matter content in the galactic halo region as perfect fluid, we
obtain space time metric at the galactic halo region in the
framework of general relativity. We find that the resultant metric
is a non-relativistic-dark-matter-induced space-time embedded in a
static Friedmann-Lema\^itre-Robertson-Walker universe. This means
that the flat rotation curve not only leads to the existence of
dark matter but also suggests the background geometry of the
universe. The flat rotation curve and the demand that the dark
matter be non-exotic together indicate a (nearly) flat universe as
favored by the modern cosmological observations. We obtain the
expressions for energy density and pressure of dark matter halo
and consequently the equation of state of dark matter. Various
other aspects of the solution are also analyzed.

\end{abstract}

\maketitle


\section{Introduction}

It has been known for a long time that rotation curves of neutral hydrogen
clouds in the outer regions of galaxies cannot be explained in terms of the
luminous matter content of the galaxies, at least within the context of
Newtonian gravity or of general relativity (see [1] for a review). In
addition, the velocity dispersion in galaxy clusters indicates a much greater
mass than would be inferred from luminous matter contained in the individual
galaxies. The most widely accepted explanation, based on the standard
gravitational theory, postulates that almost every galaxy hosts a large amount
of nonluminous matter, the so called gravitational dark matter, consisting of
unknown particles not included in the particle standard model, forming a halo
around the galaxy. The dark matter provides the needed gravitational field and
the mass required to match the observed galactic flat rotation curves in
galaxy clusters.

The dark matter problem arises because of a na\"{\i}ve Newtonian analysis of
the near constant tangential velocity of rotation up to distances far beyond
the luminous radius of the galaxies. This leads to the conclusion that the
energy density decreases with the distance as $r^{-2}$ and therefore that the
mass of galaxies increases as $m(r)\propto r$.

There are several proposals for the dark matter component, ranging from new
exotic particles such as those predicted by supersymmetry [2] to other less
exotic candidates such as massive neutrinos or even ordinary celestial bodies
such as Jupiter like objects. Several other analytic halo models exist in the
literature including those sourced by scalar fields. Fay [3] considered
modelling by Brans-Dicke massless scalar field while Matos, Guzm\'{a}n and
Nu\~{n}ez [4] considered massless minimally coupled scalar field with a
potential. A boson star formed by a self interacting massive scalar field with
quartic interaction potential as a model of galactic halo was investigated by
Colpi, Shapiro and Wasserman [5]. A similar boson star as a model of galactic
halo was first investigated by Lee and Koh [6]. A recent halo model is
discussed also in the braneworld theory [7]. Some authors [8] have considered
global monopoles as a candidate for galactic halo in the framework of the
scalar tensor theory of gravity.

Alternatively, explanation of the observed flat rotation curve is provided
through modification of Newtonian dynamics (in the region of very small
accelerations) [9,10]. Another attempt to resolve the dark matter problem lies
within the framework of Weyl conformal gravity suggested by Mannheim and
Kazanas [11], with the distinction that it does not postulate flat rotation
curve as an input but \textit{predict} it. Such a prediction is possible also
within the Chern Simmons gravity inspired by string theory [12,13].

In recent years, the amount of dark matter in the Universe has become known
more precisely: CMB anisotropy data indicate that about $85\%$ of total matter
in a galaxy is dark in nature. Big bang nucleosynthesis and some other
cosmological observations require that the bulk of dark matter be
non-baryonic, cold or warm, stable or long-lived and not interacting with
visible matter. However, despite long and intensive investigations, little is
as yet known about the nature of dark matter.

The purpose of the present work is to show that, with the input of flat
rotation curve and the assumption that dark matter be described as a perfect
fluid, the general theory of relativity not only accounts for the dark matter
component of galactic clusters but also suggests the spatial curvature of the
universe. The gravitational influence of an arbitrary dark matter component is
controlled by its stress tensor. In this context we note that, while the
existing approaches of explaining flat rotation curves are useful in their own
right, many of the models end up with predicting anisotropic dark matter fluid
stress tensor. On the other hand, no physical mechanism is stated explaining
why a spherical distribution should have such anisotropy. There is also no
observational evidence in support of it. Therefore, it seems more reasonable
to consider an isotropic perfect fluid distribution for dark matter because
predictions from such model at stellar and cosmic scales have been
observationally corroborated beyond any doubt.

We shall particularly study the general features of dark matter such as its
equation of state. The mass distribution in galactic haloes should provide a
direct probe into the nature of particles constituting the dark matter because
the inner structure of the halo is particularly sensitive to the dark matter
properties [14]. In this investigation, we shall work only from a fluid
perspective leaving the question about the particle identity of dark matter open.

\section{Gravitational field in the dark matter region}

Galactic rotational velocity profiles [15] of almost all the spiral galaxies
are characterized by a rapid increase from the galactic center, reaching a
nearly constant velocity from the nearby region of the galaxy far out to the
halo region. Our target is to exploit this observed feature to obtain the
space-time metric in the halo region, and analyze it. As mentioned already,
observational data suggest that the dark matter component in the galaxy
accounts for almost $85\%$ of its total mass. Naturally, luminous matter does
not contribute significantly to the total energy density of the galaxy,
particularly in the halo region. Therefore, we shall treat the matter in the
galactic halo region as a perfect fluid defined by stresses $T_{r}%
^{r}=T_{\theta}^{\theta}=T_{\phi}^{\phi}=p$, where $T_{\nu}^{\mu}$ is the
matter energy momentum tensor.

The general static spherically symmetric spacetime is represented by the
following metric%

\begin{equation}
ds^{2}=-e^{\nu(r)}dt^{2}+e^{\lambda(r)}dr^{2}+r^{2}(d\theta^{2}+sin^{2}\theta
d\phi^{2}),
\end{equation}
where the functions$\ \nu(r)$ and $\lambda(r)$ are the metric potentials. Then
the Einstein field equations become ($c=1$):%

\begin{equation}
e^{-\lambda}\left[  \frac{\lambda^{\prime}}{r}-\frac{1}{r^{2}}\right]
+\frac{1}{r^{2}}=8\pi G\rho
\end{equation}

\begin{equation}
e^{-\lambda}\left[  \frac{1}{r^{2}}+\frac{\nu^{\prime}}{r}\right]  -\frac
{1}{r^{2}}=8\pi Gp
\end{equation}

\begin{equation}
\frac{1}{2}e^{-\lambda}\left[  \frac{1}{2}(\nu^{\prime})^{2}+\nu^{\prime
\prime}-\frac{1}{2}\lambda^{\prime}\nu^{\prime}+\frac{1}{r}({\nu^{\prime
}-\lambda^{\prime}})\right]  =8\pi Gp.
\end{equation}
For a circular stable geodesic motion in the equatorial plane, the
consideration of flat rotation curve gives the condition (see Appendix)%

\begin{equation}
e^{\nu}=B_{0}r^{l},
\end{equation}
where $l$ is given by $l=2(v^{\phi})^{2}$ and $B_{0}$ is an integration
constant. The observed rotational curve profile in the region dominated by
dark matter is such that the rotational velocity $v^{\phi}$ becomes
approximately a constant with $v^{\phi}\sim10^{-3}$ ( 300km/s) for a typical
galaxy. The Eqs. (3) and (4) then lead to the following equation%

\begin{equation}
(e^{-\lambda})^{\prime}+\frac{ae^{-\lambda}}{r}=\frac{c}{r},
\end{equation}
where%

\begin{equation}
a=-\frac{4(1+l)-l^{2}}{2+l}%
\end{equation}
and%

\begin{equation}
c=-\frac{4}{2+l}.
\end{equation}
An exact solution of Eq.(6) is given by
\begin{equation}
e^{-\lambda}=\frac{c}{a}+\frac{D}{r^{a}},
\end{equation}
where $D$ is an integration constant. Several observations follow from the
metric (1):

\textbf{(a)} Note that the space time metric given by Eq.(1) through Eqs. (5)
and (9) is an interior solution. This type of spacetime definitely cannot be
asymptotically flat neither can it have the form of a spacetime due to a
centrally symmetric black hole. What can be said is that this line element
describes the region where the tangential velocity of the test particles is
constant and that it has to be joined with the exterior region with other
types of space time. The extent of $v^{\phi}$= constant ends at some larger
distance where the region becomes asymptotically flat. In principle, the
constant $D$ should be obtained from the junction conditions but the galactic
boundary is not observationally defined yet. It is more likely that dark
matter distribution will not remain the same near the periphery of the halo.

\textbf{(b)} Inserting the above solution in Eqs. (2) - (4), one can readily
get the expressions for $\rho$, $p$ as%

\begin{equation}
\rho= \frac{1}{8 \pi G } \left[  \frac{l(4-l)}{4+4l-l^{2}} r^{-2} -
\frac{D(6-l)(1+l)}{2+l} r^{l(2-l)/(2+l)}\right]
\end{equation}

\begin{equation}
p=\frac{1}{8\pi G}\left[  \frac{l^{2}}{4+4l-l^{2}}r^{-2}%
+D(1+l)r^{l(2-l)/(2+l)}\right]
\end{equation}

The most notable aspect is the presence of the last term in the expressions of
both energy density $\rho$ and pressure $p$ that increases with radial
distance (from galactic center). Note that according to the Newtonian theory,
which is supposed to be indistinguishable from general relativity in very weak
field, one should expect only the term $\rho_{\text{Newton}}=\frac{1}{8\pi
G}\frac{l}{r^{2}}$ in agreement with the Poisson equation, which integrates to
give the Newtonian mass $M(r)$ increasing linearly with $r$. But the input of
the constant tangential velocity leads to general relativistic corrections to
Newtonian expressions as evident in Eqs. (10), (11). The first term of the
right hand side of Eq.(10) gives the expected Newtonian term in the leading
order and to the same order a general relativistic correction term $\frac
{1}{8\pi G}\frac{5l^{2}}{4r^{2}}$. Since $l$ is small, the correction terms
are small as expected.

However, the second term in Eq.(10) has completely different nature than the
conventional feature of dark matter energy density; the corresponding mass
increases \textit{nonlinearly }with radial distance $r$, though a bit slowly
as $l$ is very small. This contribution will vanish if $D$ vanishes. We would
see that even $D=0$ is consistent with the flat rotation curve.

\textbf{(c)} The parameter $D$ is recognized [17] as the spatial curvature
(with a negative sign) of the universe. A comparison of the obtained space
time metric in the limit $l\rightarrow0$ with the static
Friedmann--Lem\^{a}itre-Robertson--Walker (FLRW) metric [ $ds^{2}%
=-dt^{2}+\frac{1}{1-kr^{2}}dr^{2}+r^{2}\left(  d\theta^{2}+sin^{2}\theta
d\phi^{2}\right)  $ ] leads to the identification that $D=-k$. Since our input
was only the flat rotation curve and we have not considered anything regarding
cosmological spatial curvature while deriving the metric, the appearance of
$D$ in the metric element is quite interesting. The solution thus may be
thought of dark matter induced space-time embedded in a static FLRW metric. In
general $l\neq0$ and hence $a$ is not equal to $-2$, and the interpretation of
spatial curvature does not seem at all evident.

An immediate question is how a spatially curved FLRW universe could be static.
If it is due to balance of the curvature term by some other fluid with
equation of state $\rho=-3p$, the question of stability will arise. It seems
that we have ended up with a static universe because right from the beginning
we looked for a static solution (all the metric element are considered time
independent). While working on a local problem (flat rotation curve), the
universe is usually considered at any particular epoch fixing the scale factor
$R(t)$ to be constant (often normalized to unity, $R(t_{0})=1$ at the present
epoch). The dynamicity of the FLRW metric is thus absent in local
gravitational phenomena. However, we see in the present work that the
curvature effect has appeared even in local gravitational phenomena. It would
certainly be an interesting but challenging task to study the stability
criterion in this configuration [17]. We reserve it for a separate communication.

\textbf{(d)} The equation of state parameter $\omega$ for the effective fluid
(dark matter plus \lq curvature fluid \rq), can be obtained directly from
Eqs.(10) and (11), which is given by%

\begin{equation}
\frac{p}{\rho}=\omega=\frac{l^{2}r^{a}+D(1+l)(4+4l-l^{2})}{l(4-l)r^{a}%
-D(6-l)(1+l)(4+4l-l^{2})/(2+l)}.
\end{equation}
Since $l$ is small, effectively $\omega\simeq\frac{l^{2}r^{a}+4D}{4lr^{a}%
-12D}$. If the total matter in the flat rotation curve region of galaxies has
to be non-exotic, i.e. if the dark matter satisfies the known energy
conditions, $\omega$ for the total matter content must be positive. This is
achieved only if we set $l^{2}r^{a}/4<D<lr^{a}/3$. Note that our discussion is
restricted only in the region of flat rotation curve of galaxies for which $r$
is typically between few ten kpc to few hundred kpc and as mentioned already
$v^{\phi}$ is few hundred km/s. Hence positive $\omega$ implies that $D$ is
nearly zero, if not exactly. Figure 1 shows that variation of $\omega$ with
$D$ for a typical distance $r=200\;kpc$ and $v^{\phi}=300\;km/s$.

\begin{figure}[ptb]
\begin{center}
\vspace{0.5cm} \includegraphics[width=0.5\textwidth]{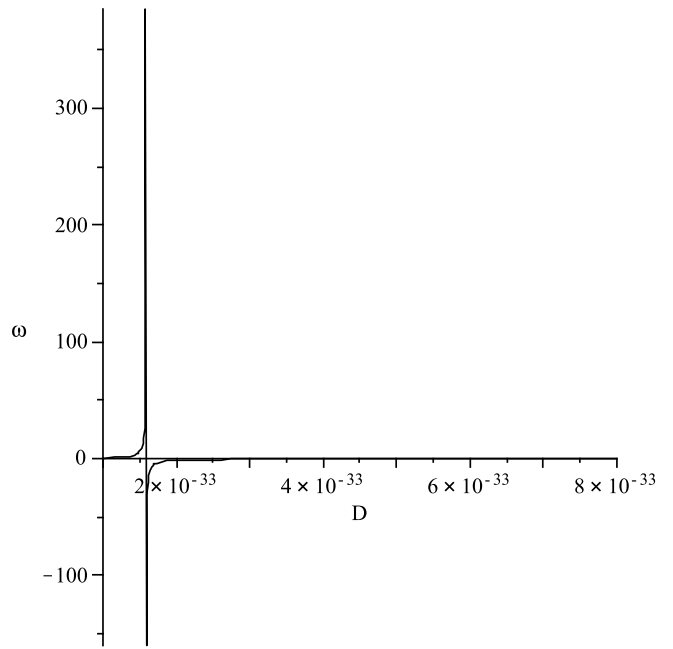}
\end{center}
\caption{The variation of $\omega$ with $D$ for a typical distance $r=200$ in
$Kpc$.}%
\label{fig3}%
\end{figure}

As may be seen from the figure 1, the limit on $D$ is much stringent compared
to the (cosmological) observational restriction. The equation of state of the
dark matter component may be obtained by taking $D=0$ in the above equation
which gives%

\begin{equation}
\omega=\frac{l}{(4-l)}\simeq2.5\times10^{-7}\text{. }%
\end{equation}
Thus $p<<\rho$ for dark matter implying its non-relativistic nature, which is
a well known fact. Further $\omega$ is positive, which means that the fluid is
non-exotic. On the other hand, the equation of state of `curvature fluid 'is
found to be $\rho+3p=0$ from Eq.(12) in the limit $l\rightarrow0$, which is a
familiar result.

\textbf{(e)} Note that we can rewrite $e^{\lambda}$ \ in the standard
Schwarzschild form%

\begin{equation}
e^{\lambda}=\left[  1-\frac{2m(r)}{r}\right]  ^{-1}%
\end{equation}
which is often convenient. Such a form has the advantage that it immediately
reveals not only the mass parameter $m(r)$ but also shows that the proper
radial length is larger than the Euclidean length because $r>2m(r)$. But most
importantly, this inequality is essential for signature protection, which
dictates that $e^{\lambda}>1$. This is a crucial condition to be satisfied by
any valid metric. Now, from the metric function (9), for $D=0$, we get%
\begin{equation}
e^{\lambda}=1+\frac{4l-l^{2}}{4}>1.
\end{equation}
This implies that the essential requirement is fulfilled for this value of $D
$.

\textbf{(f)} Let us rewrite the metric (1) for $D=0$ under the radial
rescaling%
\begin{equation}
r=\sqrt{\frac{c}{a}}r^{\prime},
\end{equation}
which yields%
\[
ds^{2}=-B_{0}^{\prime}r^{\prime l}dt^{2}+dr^{\prime2}+\left(  \frac{c}%
{a}\right)  r^{\prime2}(d\theta^{2}+sin^{2}\theta d\phi^{2}),\newline%
\]
\begin{equation}
\text{\ }B_{0}^{\prime}=B_{0}\left(  \sqrt{\frac{c}{a}}\right)  ^{l}%
\end{equation}
showing a surplus angle in the surface area given by
\begin{equation}
S_{1}=4\pi r^{2}.\frac{c}{a}=\frac{4\pi r^{2}}{1+2(v^{\phi})^{2}-(v^{\phi
})^{4}}.
\end{equation}
If the probe particles were photons, so that $v^{\phi}=1$, the surface area
would remain \textit{finite} but reduced to half of the spherical surface area
$S_{2}=4\pi r^{2}$. This is an interesting result, which distinguishes itself
from that in the massless scalar field model where $S\rightarrow\infty$ as
$v^{\phi}=1$ [4]. For a typical rotational velocity of $v^{\phi}=10^{-3}$ in
the galactic halo region, the difference of the two surface areas
\begin{equation}
S_{2}-S_{1}=4\pi r^{2}\left[  \frac{2(v^{\phi})^{2}-(v^{\phi})^{4}%
}{1+2(v^{\phi})^{2}-(v^{\phi})^{4}}\right]
\end{equation}
grows as $\sim10^{-6}$ in units of flat surface area, which indicates another
deviation from the massless scalar model in which it grows as $\sim10^{-12}$ [4].

\textbf{(g)} The Ricci scalar for the derived spacetime is given by
\begin{equation}
R=\frac{Da^{2}(4+l)-(aD+cr^{a})(l^{2}+2l+4)+4ar^{a}}{2ar^{2+a}}%
\end{equation}
As $l\rightarrow0$, $R=-6D$, once again suggesting that $D$ is the spatial
curvature. We plot $R$ vs $r$. One may note that the value of $R$ is small.
For small different values of $D$, one can not distinguish the variations of
$R$ with respect to $r$ ( see figure 2 ). But for higher values of $D$, figure
3 shows a sharp variation of $R$. Note that large values of $D$ imply
$\omega(r)<0$, i.e., dark matter has to be exotic in nature.

\begin{figure}[ptb]
\begin{center}
\vspace{0.5cm} \includegraphics[width=0.4\textwidth]{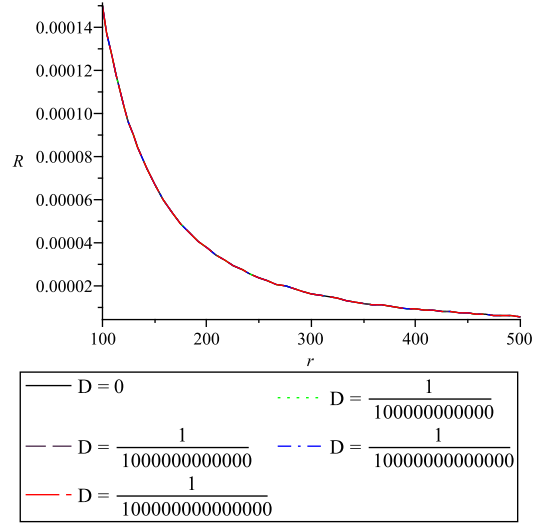}
\end{center}
\caption{The variation of $R$ with $r$ in $Kpc$ for small different values of
D. We choose $v^{\phi}\sim10^{-3}$ (300km/s) for a typical galaxy.}%
\label{fig3}%
\end{figure}\begin{figure}[ptb]
\begin{center}
\vspace{0.5cm} \includegraphics[width=0.4\textwidth]{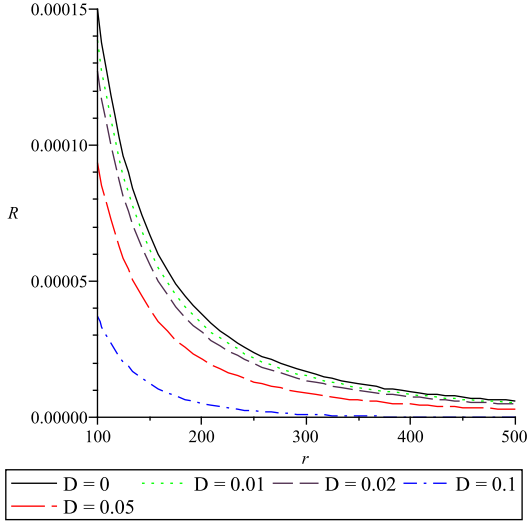}
\end{center}
\caption{The variation of $R$ with $r$ in $Kpc$ for large different values of
D. We choose $v^{\phi}\sim10^{-3}$ (300km/s) for a typical galaxy. }%
\label{fig3}%
\end{figure}

\section{Other aspects}

\subsection{ \underline{\textbf{\ The total gravitational energy}}:}

One notes from equation (12) that the halo matter is not exotic in nature and
consequently, we expect attractive gravity in the halo. Following the
suggestion given by Lyndell - Bell et al [18] , we calculate the total
gravitational energy $E_{G}$ between two fixed radii , say, $r_{1}$ and
$r_{2}$:
\begin{align}
E_{G}  & =M-E_{M}=4\pi\int_{r_{1}}^{r_{2}}[1-\sqrt{e^{\lambda(r)}}]\rho
r^{2}dr\newline\\
& =4\pi\int_{r_{1}}^{r_{2}}\left[  1-\sqrt{\frac{1}{\frac{c}{a}+\frac{D}%
{r^{a}}}}\right]  \left[  \frac{1}{8\pi G}\left(  \frac{D(a-1)}{r^{a+2}}%
+\frac{(1-\frac{c}{a})}{r^{2}}\right)  \right]  r^{2}dr,
\end{align}
where
\begin{equation}
M=4\pi\int_{r_{1}}^{r_{2}}\rho r^{2}dr
\end{equation}
is the Newtonian mass given by%

\begin{equation}
M=4\pi\int_{r_{1}}^{r_{2}}\rho r^{2}dr=\frac{1}{G}\left[  \frac{(1-\frac{c}%
{a})r}{2}-\frac{D}{2r^{a-1}}\right]  _{r_{1}}^{r_{2}}.
\end{equation}
Thus we get the total gravitational energy as
\begin{align}
E_{G}  & =\frac{1}{G}\left[  \frac{(1-\frac{c}{a})r}{2}-\frac{D}{2r^{a-1}%
}-(1-\frac{c}{a})\frac{rF[(0.5,\frac{1}{a});(1+\frac{1}{a});-\frac{ar^{a}D}%
{c}]}{\sqrt{\frac{c}{a}}}\right]  _{r_{1}}^{r_{2}}\newline\nonumber\\
& +\frac{1}{G}\left[  D(1-a)r^{(-0.5a+1)}\frac{F[(-0.5+\frac{1}{a}%
,0.5);(0.5+\frac{1}{a});-\frac{cr^{a}}{aD}]}{\sqrt{D}a(-0.5+\frac{1}{a}%
)}\right]  _{r_{1}}^{r_{2}}.
\end{align}

The figures 4 and 5 show that the total gravitational energy is small but
negative whether we choose $D$ non-zero or zero for arbitrary $r_{2}>r_{1}>0$.
Thus for the existence of non-exotic matter in the halo, we recommend the
value of $D\leq10^{-11}$. In the distant halo region we have taken, typically
, $r$ $\sim200$ kpc in the figures below.

\begin{figure}[ptb]
\begin{center}
\vspace{0.5cm}\includegraphics[width=0.5\textwidth]{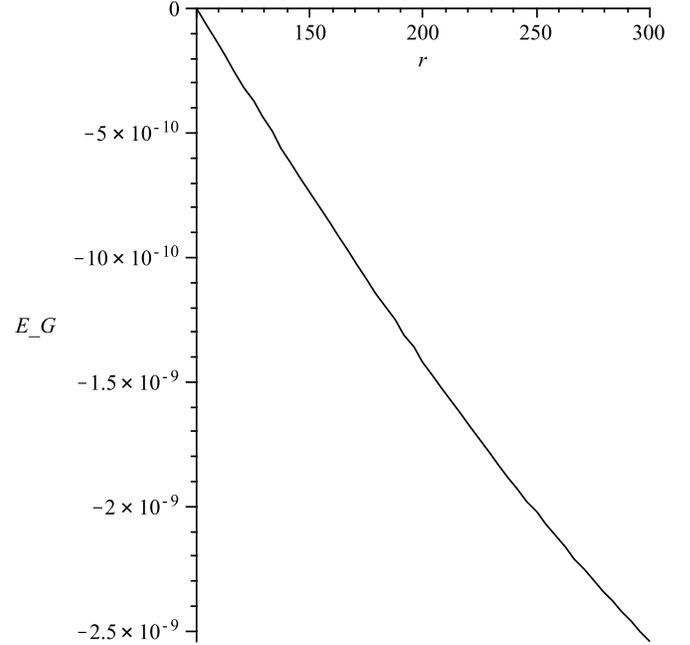}
\end{center}
\caption{The variation of $E_{G}$ with ~r~ in Kpc. The lower limit of
integration in equation (24) fixed at, say, $r_{1} = 100$ kpc while $r_{2}$ is
varied from 100 to 500 Kpc. We choose $G=1$, $D=10^{-11}$ and $v^{\phi}%
\sim10^{-3}$ ( 300km/s) for a typical galaxy.}%
\label{fig3}%
\end{figure}

\begin{figure}[ptb]
\begin{center}
\vspace{0.5cm}\includegraphics[width=0.4\textwidth]{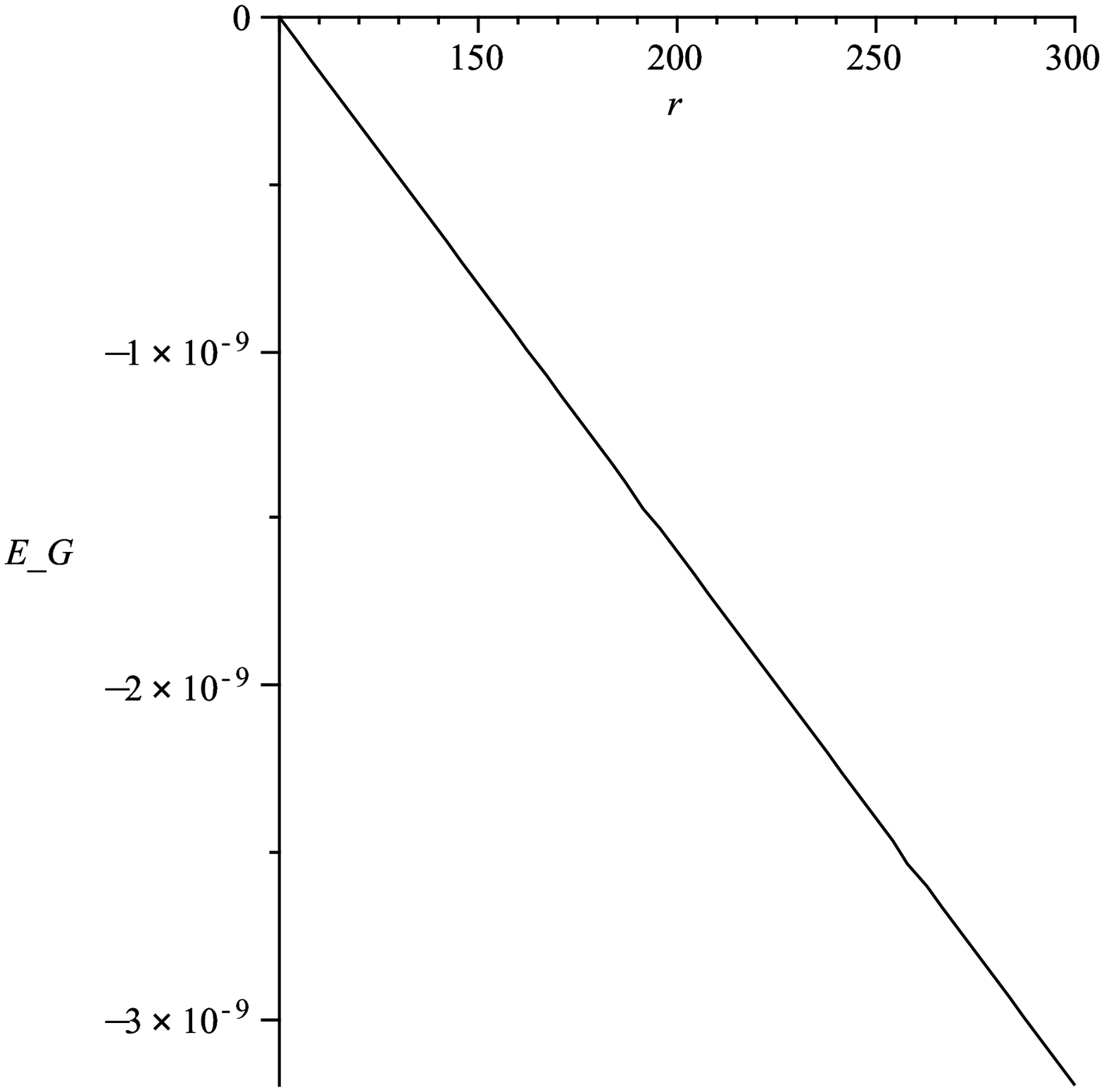}
\end{center}
\caption{The variation of $E_{G}$ with~ r ~in Kpc. The lower limit of
integration in equation (24) fixed at, say, $r_{1} = 100$ kpc while $r_{2}$ is
varied from 100 to 500 Kpc. We choose $G=1$ , $D=0$ and $v^{\phi}\sim10^{-3}$
( 300km/s) for a typical galaxy.}%
\label{fig3}%
\end{figure}

\subsection{ \underline{\textbf{\ Attraction}}:}

Now we study the geodesic equation given by
\begin{equation}
\frac{d^{2}x^{\alpha}}{d\tau^{2}}+\Gamma_{\alpha}^{\mu\gamma}\frac{dx^{\mu}%
}{d\tau}\frac{dx^{\gamma}}{d\tau}=0
\end{equation}
for a test particle that has been \textquotedblleft placed'\ at some radius
$r_{0}$. This yields the radial equation%

\begin{eqnarray}
\frac{d^{2}r}{d\tau^{2}}&=&-\frac{1}{2}\left[  \frac{c}{a}+\frac{D}{r^{a}%
}\right]  \nonumber \\
&& \left[  \frac{Da}{r^{a+1}}\left(  \frac{c}{a}+\frac{D}{r^{a}%
}\right)  ^{-2}\left(  \frac{dr}{d\tau}\right)  ^{2}+B_{0}lr^{l-1}\left(
\frac{dt}{d\tau}\right)  ^{2}\right]  _{r=r_{0}},
\end{eqnarray}
\newline which is negative as the the quantity in the square bracket is
positive. Thus particles are attracted towards the center. This result is in
agreement with the observations i.e., gravity on the galactic scale is
attractive ( clustering , structure formation etc ).

\subsection{ \underline{\textbf{\ Stability:}}}

Let us define the four velocity $U^{\alpha}=\frac{dx^{\sigma}}{d\tau}$ for a
test particle moving solely in the space of the halo (restricting ourselves to
$\theta=\pi/2$), the equation $g_{\nu\sigma}U^{\nu}U^{\sigma}=-m_{0}^{2}$ can
be cast in a Newtonian form
\begin{equation}
\left(  \frac{dr}{d\tau}\right)  ^{2}=E^{2}+V(r)
\end{equation}
which gives%
\begin{equation}
V(r)=-\left[  E^{2}\left\{  1-\frac{r^{-l}\left[  \frac{c}{a}+\frac{D}{r^{a}%
}\right]  }{B_{0}}\right\}  +\left[  \frac{c}{a}+\frac{D}{r^{a}}\right]
\left(  1+\frac{L^{2}}{r^{2}}\right)  \right]
\end{equation}%
\begin{equation}
E=\frac{U_{0}}{m_{0}},L=\frac{U_{3}}{m_{0}},
\end{equation}
where the constants $E$ and $L$, respectively, are the conserved relativistic
energy and angular momentum per unit rest mass of the test particle. Circular
orbits are defined by $r=R=$constant. so that $\frac{dR}{d\tau}=0$ and,
additionally, $\frac{dV}{dr}\mid_{r=R}=0$. From these two conditions follow
the conserved parameters:
\begin{equation}
L=\pm\sqrt{\frac{l}{2-l}}R
\end{equation}
and using it in $V(R)=-E^{2}$, we get
\begin{equation}
E=\pm\sqrt{\frac{2B_{0}}{2-l}}R^{\frac{l}{2}}.
\end{equation}
The orbits will be stable if $\frac{d^{2}V}{dr^{2}}\mid_{r=R}<0$ and unstable
if $\frac{d^{2}V}{dr^{2}}\mid_{r=R}>0$. Putting the expressions for $L$ and
$E$ in $\frac{d^{2}V}{dr^{2}}\mid_{r=R}$, we obtain, after straightforward
calculations, the final result, viz.,
\begin{equation}
\frac{d^{2}V}{dr^{2}}\mid_{r=R}=-\left[\frac{2c l (1-2l)}{a(2-l) R^2}+\frac{%
D( 16 l +8l^2-4l^3+2l^4) }{(2-l)(2+l)^2}
R^\frac{l(2-l)}{2+l}\right]
\end{equation}
One may note that, for $D=0$ as well as for $D\neq0$, $\frac{d^{2}V}{dr^{2}%
}\mid_{r=R}<0$. So the circular orbits are always stable.

\section{ Conclusions:}

It is well known that observation of flat rotation curve suggests that a
substantial amount of non-luminous dark matter is hidden in the galactic halo.
Here we have found that the flat rotation curve suggests also the background
geometry of the universe. The space-time geometry we have obtained can be
interpreted as the one due to dark matter embedded in the static FLRW
universe. This is probably the first indication that the spatial curvature of
the universe can be obtained from a local gravitational phenomenon.

If we demand that matter in the flat rotation curve region be non-exotic
(i.e., obey the usual energy conditions), we obtain the result that the
universe should be nearly flat, if not exactly so, which is consistent with
modern cosmological observations.

The equation of state of the dark matter component has been obtained by
treating it as perfect fluid and the expressions for general relativistic
correction terms for the pressure and energy density over those obtained from
the Newtonian theory are also derived. The corrections are, however, small as expected.

The geodesic equation [Eq.(26)] of a test particle in the derived spacetime
suggests that the particle will be attracted towards the center. We have
quantified the attractive effect in the relativistic case by calculating the
total gravitational energy $E_{G}$ ( which is negative ) in the halo region.
We have also demonstrated the stability of the circular orbits in our
spacetime solution. Thus our solution satisfies two crucial physical
requirements - stability of circular orbits and attractive gravity in the halo
region. Investigation on other observational constraints on the model is underway.

\subsection*{Acknowledgments}

The authors are thankful to an anonymous referee for his/her insightful
comments that have led to significant improvements, particularly on the
interpretational aspects. FR is grateful to IMSc, Chennai and IUCAA, Pune for
providing research facilities. \newline

\subsection*{Appendix}

To derive tangential velocity of circular orbits, we start with the line
element
\[
ds^{2}=-e^{\nu(r)}dt^{2}+e^{\lambda(r)}dr^{2}+r^{2}(d\theta^{2}+sin^{2}\theta
d\phi^{2}).
\]
The Lagrangian for a test particle reads
\[
2\mathbf{\mathit{L}}=-e^{\nu(r)}\dot{t}^{2}+e^{\lambda(r)}\dot{r}^{2}%
+r^{2}(\dot{\theta}^{2}+sin^{2}\theta\dot{\phi}^{2}).
\]
We guess the conserved quantities, the energy $E=e^{\nu(r)}\dot{t}$, the
$\phi$-momentum $L_{\phi}=r^{2}sin^{2}\theta\dot{\phi}$, and the total angular
momentum, $L^{2}={L_{\theta}}^{2}+\left(  \frac{L_{\phi}}{\sin\theta}\right)
^{2}$, with $L_{\theta}=r^{2}\dot{\theta}$. The radial motion equation can be
written as:
\[
\dot{r}^{2}+V(r)=0
\]
with the potential $V(r)$ given y
\[
V(r)=-e^{-\nu(r)}\left(  e^{-\lambda(r)}E^{2}-\frac{L^{2}}{r^{2}}-1\right)  .
\]
For circular orbits, we have the conditions, $\dot{r}=0$, $V_{r}=0$ and
$V_{rr}~>~0$. These imply the following expressions for the energy and total
momentum of the particles in circular orbits:
\[
E^{2}=\frac{2e^{2\nu(r)}}{2e^{\nu(r)}-r(e^{\nu(r)})_{r}}%
\]%
\[
L^{2}=\frac{r^{3}(e^{\nu(r)})_{r}}{2e^{\nu(r)}-r(e^{\nu(r)})_{r}}.
\]
The second derivative of the potential evaluated at the extrema
\[
V(r)_{rr}|_{\text{extrema}}=2\frac{\frac{r(e^{\nu(r)})_{rr}}{e^{\nu(r)}}%
+\frac{(e^{\nu(r)})_{r}}{e^{\nu(r)}}\left(  3-\frac{r(e^{\nu(r)})_{r}}%
{e^{\nu(r)}}\right)  }{re^{\lambda(r)}\left(  2-\frac{r(e^{\nu(r)})_{r}%
}{e^{\nu(r)}}\right)  }%
\]
Now, the tangential velocity
\[
(v^{\phi})^{2}=r^{2}e^{-\nu(r)}(\dot{\theta}^{2}+sin^{2}\theta\dot{\phi}^{2})
\]
can be obtained for particles in stable circular orbits as
\[
(v^{\phi})^{2}=\frac{r(e^{\nu(r)})_{r}}{2e^{\nu(r)}}.
\]
Using the tangential velocity to be constant for several radii, the above
expression yields
\[
e^{\nu}=B_{0}r^{l}%
\]
where $l$ is given by $l=2(v^{\phi})^{2}$ and $B_{0}$ is an integration constant.

\subsection*{\textbf{References}}

[1] G. Bertone, D. Hooper, and J. Silk, Phys. Rep. \textbf{405}, 279 (2005).

[2] G. Jungman, M. Kamionkowski and K. Griest, Phys. Rep. \textbf{267}, 195 (1996).

[3] S. Fay, Astron. Astrophys. \textbf{413}, 799 (2004).

[4] T. Matos, F. S. Guzm\'{a}n, and D. Nu\~{n}ez, Phys. Rev. D \textbf{62},
061301 (2000). For a further analysis of the model, see: K.K. Nandi, I.
Valitov, and N.G. Migranov, Phys. Rev. D \textbf{80}, 047301 (2009).

[5] M. Colpi, S. L. Shapiro, and I. Wasserman, Phys. Rev. Lett. \textbf{57},
2485 (1986).

[6] J.-w. Lee and I.-g. Koh, Phys. Rev. D \textbf{53}, 2236 (1996).

[7] F. Rahaman, M. Kalam, A. DeBenedictis, A. A. Usmani, and Saibal Ray, Mon.
Not. R. Astron. Soc. \textbf{389}, 27 (2008); K. K. Nandi, A.I. Filippov, F.
Rahaman, Saibal Ray, A. A. Usmani, M. Kalam, and A. DeBenedictis, Mon. Not. R.
Astron. Soc. \textbf{399}, 2079 (2009).

[8] U. Nukamendi, M. Salgado, and D. Sudarsky, Phys. Rev. Lett., \textbf{84},
3037 (2000); T. Lee and B. Lee, Phys. Rev. D \textbf{69}, 127502 (2004); F.
Rahaman, R. Mondal, M. Kalam, and B. Raychaudhuri, Mod. Phys. Lett. A
\textbf{22}, 971 (2007).

[9] M. Milgrom, Astrophys. J. \textbf{270}, 365 (1983).

[10] J.D. Bekenstein J. and M. Milgrom, Astrophys. J. \textbf{286}, 7 (1984).

[11] P. D. Mannheim and D. Kazanas, Astrophys. J. \textbf{342}, 635 (1989).

[12] K. Konno, T. Matsuyama, Y. Asano, and S. Tanda, Phys. Rev. D \textbf{78},
024037 (2008).

[13] K.K. Nandi, I.R. Kizirgulov, O.V. Mikolaychuk, N.P. Mikolaychuk, A.A.
Potapov, Phys. Rev. D \textbf{79}, 083006 (2009).

[14] S. Tremaine and J. Gunn, Phys. Rev. Lett. \textbf{42}, 407 (1979); G.
Lake, Astron. J, \textbf{98}, 1253 (1989).

[15] V. C. Rubin and W. K. Ford, Astrophys. J. \textbf{159}, 379 (1970); J. N.
Bahcall and S. Casertano, Astrophys. J. \textbf{293}, L7 (1985); D. Burstein
and V. C. Rubin, Astrophys. J. \textbf{297}, 423 (1985); M. Persic and P.
Salucci, Astrophys. J. Suppl. \textbf{99}, 501 (1995).

[16] S. Chandrasekhar, \textit{Mathematical Theory of Black Holes }(Oxford
University Press, Oxford, 1983).

[17] We sincerely thank the referee for raising this interesting possibility.

[18] D. Lynden-Bell, J. Katz, and J. Bi\v{c}\'{a}k J, Phys. Rev.
D\textbf{\ 75}, 024040 (2007); See for an extension: K.K. Nandi, Y.Z. Zhang,
R.G. Cai, and A. Panchenko, Phys. Rev. D\textbf{\ 79}, 024011 (2009).

\end{document}